\documentclass[journal]{IEEEtran}
\usepackage{graphicx,amssymb,lineno}
\usepackage{subfigure}
\usepackage{amsmath,amsfonts,amssymb}

\usepackage[linesnumbered,ruled,vlined]{algorithm2e}
\usepackage{algorithmic}
\usepackage[usenames]{color}
\usepackage{float}
\usepackage{multirow} 
\usepackage{amsmath}

\usepackage{times, verbatim, subfigure, epsfig, graphicx, latexsym, amsmath}
\usepackage{url}
\usepackage{subfigure}
\usepackage{CJK}

\begin{document}

\title{Relay-Assisted and QoS Aware Scheduling  \\ to Overcome Blockage in mmWave Backhaul Networks}

\author{Yong~Niu,~\IEEEmembership{Member,~IEEE,}
        Weiguang~Ding,
        Hao~Wu,~\IEEEmembership{Member,~IEEE,}
        Yong~Li,~\IEEEmembership{Senior Member,~IEEE},
        Xinlei Chen,
        Bo Ai,~\IEEEmembership{Senior Member,~IEEE}
        and Zhangdui Zhong,~\IEEEmembership{Senior Member,~IEEE}
\thanks{Y.~Niu, W.~Ding, H.~Wu, B.~Ai, and Z. Zhong are with the State Key Laboratory of Rail Traffic
Control and Safety, the School of Electronic and Information Engineering, and Beijing Engineering Research Center of High-speed Railway Broadband Mobile Communications, Beijing Jiaotong University, Beijing 100044, China (e-mails:16120056@bjtu.edu.cn; niuy11@163.com; hwu@bjtu.edu.cn).}

\thanks{Y. Li is with State Key Laboratory on Microwave and Digital
 Communications, Tsinghua National Laboratory for Information Science and Technology
 (TNLIST), Department of Electronic Engineering, Tsinghua University, Beijing 100084, China (E-mails:
  liyong07@tsinghua.edu.cn).}
 \thanks{X.~Chen is with the Department of Electrical and Computer Engineering, Carnegie Mellon University, Pittsburgh, U.S..}
\thanks{This study was supported by National Key R\&D Program of China under Grant 2016YFE0200900; and by the National Natural Science Foundation of China Grants 61725101 and 61801016; and by the China Postdoctoral Science Foundation under Grant 2017M610040 and 2018T110041; and by the Beijing Natural Fund under Grant L172020; and by Major projects of Beijing Municipal Science and Technology Commission under Grant No. Z181100003218010.}
}

\maketitle
\begin{abstract}

In the scenario where small cells are densely deployed, the millimeter wave (mmWave) wireless backhaul network has been widely used. However, mmWave is easily blocked by obstacles, and how to forward the data of the blocked flows is still a significant challenge.
To ensure backhauling capacity, the quality of service (QoS) requirements of flows should be satisfied.
In this paper, we investigate the problem of optimal scheduling to maximize the number of flows satisfying their QoS requirements with relays exploited to overcome blockage. To achieve a practical solution, we propose a relay-assisted and QoS aware scheduling scheme for the backhaul networks, called RAQS. It consists of a relay selection algorithm and a transmission scheduling algorithm.
The relay selection algorithm selects non-repeating relays with high link rates for the blocked flows, which helps to achieve the QoS requirements of flows as soon as possible. Then, according to the results of relay selection, the transmission scheduling algorithm exploits concurrent transmissions to satisfy the QoS requirements of flows as much as possible. Extensive simulations show RAQS can effectively overcome the blockage problem, and increase the number of completed flows and network throughput compared with other schemes. In particular, the impact of relay selection parameter is also investigated to further guide the relay selection.
\end{abstract}


\IEEEpeerreviewmaketitle
\section{Introduction}
With the rapid growth of mobile data demand, it's becoming a trend that densely deploying small cells underlying the homogeneous macrocells to improve network capacity. This kind of network is usually referred to as heterogeneous cellular network (HCN) \cite{Niu multihop}. Because of the huge available bandwidth in the millimeter wave (mmWave) band, such as the 60GHz band and E-band (71-76 GHz and 81-86 GHz), mmWave wireless backhaul communication can provide multi-gigabit transmission rates and support a lot of high-speed data services. Compared with the fibre-based backhaul communication, it's more cost-effective, more flexible and easier to deploy. Thus, it has become a candidate solution for the fifth generation (5G) mobile communication.

Compared with other electromagnetic waves at lower frequencies, mmWave communication has three main characteristics: high propagation loss, directivity, and vulnerability to obstacles \cite{survey,channel-he,Jiayi}. We usually adopt directional antennas to combat the high propagation loss. The beamforming techniques are used to direct the beams of the transmitter and receiver towards each other. Under the directional communication, the multi-user interference (MUI) between different links is reduced, and thus concurrent transmissions (i.e. spatial reuse) can be fully utilized to improve the transmission efficiency and increase the network capacity.

\textbf{Motivation:}
However, because of the vulnerability to obstacles, the flows (i.e. the traffic data transmitted between two stations) in mmWave band are easy to be blocked, which seriously affect the user's experience for delay-sensitive applications, e.g., high-definition television (HDTV). For the blocked flows, how to ensure the data transmission has become an urgent problem to be solved.
Besides, some bandwidth-intensive applications supported by mmWave networks, such as uncompressed video streaming, should be provided with multi-Gbps throughput to guarantee the transmission quality \cite{Qiao2}. Therefore, in order to ensure backhauling capacity, the quality of service (QoS) requirements of flows should also be taken into account. Here, the QoS requirements means the minimum throughput requirements.

\textbf{Main Contributions:} According to the above analysis, in this paper, we aim at solving the blockage problem and maximizing the number of flows satisfying their QoS requirements. We propose a relay-assisted and QoS aware (RAQS) scheduling scheme to overcome blockage. In the scheme, we consider to actively deploy relays in the small cells densely deployment scenario to forward data for the blocked flows in the backhaul network. The relay node has simple structure and is easy to be deployed. It has lower cost and is more flexible compared with base station (BS). Furthermore, using relay can reduce the traffic load of BS and the complexity of scheduling.

In RAQS, we optimize the relay selection for the blocked flows. Since the number of slots in one superframe is limited and all nodes are assumed to be half-duplex, not any relay is beneficial to achieve the flows' QoS requirements. Therefore, the relay path should have a high rate so that the demand of one flow can be completed as soon as possible and more slots can be saved to let other flows transmit. Besides, different blocked flows should choose different relays to reduce the node contention and allow more flows to transmit concurrently, which can achieve more flows' QoS requirements. After establishing the relay path, an efficient and low-complexity scheduling algorithm when the relay paths and backhaul paths coexist is proposed. It fully exploits concurrent transmissions to satisfy the QoS requirements of flows. The contributions of this paper are summarized as follows.

\begin{itemize}

\item We formulate the optimal concurrent transmission scheduling problem of the mmWave backhaul network with the relay paths and backhaul paths considered into a mixed integer nonlinear programming (MINLP) problem. In order to ensure backhauling capacity and guarantee fairness, we aim at maximizing the number of flows with their QoS requirements satisfied.
\item We design a relay selection algorithm to select appropriate relay(s) for the blocked flow(s). The rate of the selected relay paths are high enough and different blocked flows select different relays. In this way, we can make full use of concurrent transmissions to achieve the QoS requirements of flows in the limited time of one superframe.
\item To achieve a practical solution, we propose an heuristic algorithm to solve the joint scheduling problem of relay paths and backhaul paths. The interference between concurrent flows and the difference between the two-hop relay path and the one-hop backhaul path are considered to meet the QoS requirements of more flows and improve the network throughput.
\item We conduct extensive simulations in the mmWave band to evaluate the performance of our RAQS scheme. The results demonstrate our scheme can guarantee the number of completed flows and the system throughput at a high and stable level. Particularly, we also investigate the impact of the relay selection parameter on system performance.
\end{itemize}

The rest of this paper is organized as follows. Section \ref{S2} introduces the related work. Section \ref{S3} introduces the system model and assumption. In Section \ref{S4}, we formulate the optimal scheduling problem when relay paths and backhaul paths coexist into an MINLP, and then in Section \ref{S5}, the relay selection algorithm and corresponding scheduling algorithm in RAQS are described in detail. In Section \ref{S5-b}, we analyze the impact of interference threshold choice on the performance of our scheme.
Finally, we present the simulation results in Section \ref{S6} and we conclude this paper in Section \ref{S7}.

\section{Related Work}\label{S2}
Recently, the mmWave network in the scenario where small cells are densely deployed has gained much attention. Taori \emph{et al.} \cite{Taori} considered the time-division multiplexing (TDM)-based scheduling scheme for the backhaul network, but the directivity of mmWave and concurrent transmissions are not exploited. Qiao \emph{et al.} \cite{Qiao1} proposed a slot resource sharing scheme in the mmWave 5G cellular network, where D2D
communications and concurrent transmissions are enabled to improve network capacity. However, to simplify the problem, only non-interfering links are allocated to each time slot to share the resources \cite{Qiao1}. Later, Qiao \emph{et al.} \cite{Qiao2} proposed a STDMA-based scheme in mmWave WPAN, where both non-interfering and interfering links are allowed to be transmitted concurrently.
With the QoS requirements of flows considered, the main idea of the scheme is that if scheduling one flow can increase the system throughput, we then decide to schedule it. In this way, the number of flows satisfying their QoS requirements is also maximized. Based on \cite{Qiao2}, Zhu \emph{et al.} \cite{zhuyun} proposed a maximum QoS-aware independent set scheduling algorithm named MQIS in the mmWave backhaul network. In MQIS, the QoS aware priority is exploited to further increase the number of flows satisfying their QoS requirements and the system throughput. In \cite{NiuD2D}, a joint transmission scheduling algorithm for the radio access and backhaul of small cells in the mmWave band was proposed. However, all schemes mentioned above (\cite{Taori}, \cite{Qiao2}, \cite{zhuyun} and \cite{NiuD2D}) don't consider the scenario where the flows may be blocked. In \cite{Niu-tcom}, both D2D communications and concurrent transmissions are exploited to improve the energy efficiency of multicast transmission in mmWave small cells, where power control is performed after concurrent transmission and D2D transmission scheduling to reduce energy consumption with the achieved throughput ensured. However, the blockage problem is not considered.
In \cite{liuyu-twc}, multi-hop D2D transmissions are exploited to optimize the transmission scheduling from the base station to the service points, where the mobility information is considered.
In \cite{Ding-access}, full duplex mmWave communication is utilized to achieve better quality of service guarantee in terms of throughput for flows in mmWave backhaul networks.
In \cite{CONMD2D}, concurrent transmissions, the multi-level antenna codebook, and D2D communications are jointly exploited to improve the throughput of multicast transmissions in mmWave small cells.

There are also some literatures focusing on the flow blockage problem. Genc \emph{et al.} \cite{reflection} tried to rely on the reflections from walls and other surfaces to overcome the obstruction. Singh \emph{et al.} \cite{static reflection} used strategically placed reflectors to provide alternate paths for the blocked paths. Nevertheless, in these schemes, the power efficiency is reduced because of the power loss on the reflective surface and the extra path loss caused by longer transmission path.
In \cite{switch}, the authors resolved flow blockage by switching the beam path from a LOS link to a NLOS link, but NLOS transmissions will suffer from huge attenuation compared with LOS transmissions.
In \cite{beamswitching}, an analog beam selection scheme with low complexity is proposed, and furthermore, a beam switching scheme based on channel state information is proposed to overcome blockage problem.

In \cite{Qiao1}, Qiao \emph{et al.} proposed a relaying mechanism to reduce the link outage probability by replacing a blocked link with an alternative path. Niu \emph{et al.} \cite{Niu multihop} and Qiao \emph{et al.} \cite{Qiao multihop} used other non-PNC (piconet controller) stations in WPAN to establish multi-hop paths to overcome blockage. However, Qiao \emph{et al.} \cite{Qiao multihop} more focus on how to exploit multiple short hops to improve the flow throughput and balance the traffic loads across the network. Singh \emph{et al.} \cite{relay2} proposed a novel multihop medium access control (MAC) architecture for the 60 GHz in-room WPAN. In this architecture, if the LOS path between the access point (AP) and the wireless terminal (WT) is obstructed, the AP intelligently chooses a WT in the neighboring sectors (with expected LOS connectivity to the lost WT) to act as a relay for future data transfers. However, it doesn't consider the QoS requirements of flows. In \cite{relay1}, Leong \emph{et al.} proposed a 3D pyramid network infrastructure consisting of a single AP with four (but not restricted to) active relays operating in parallel to overcome the obstructions, but they also don't consider the QoS requirements of flows and don't talk about the specific relay selection algorithm. Resulting from the limited slot resources and the half-duplex nature of nodes, we must note that not any relay is beneficial to satisfy the QoS requirements of flows.

In this paper, we first develop a relay-assisted and QoS aware scheduling (RAQS) scheme, which exploits independent relay nodes to solve the blockage problem in mmWave backhaul network, and considers the QoS requirements of flows at the same time. Specifically, it consists of a relay selection algorithm and a transmission scheduling algorithm when relay paths and backhaul paths coexist.

\section{System Model and Assumption}\label{S3}
\begin{figure}[htbp]
  \begin{center}
  \includegraphics[width=9cm]{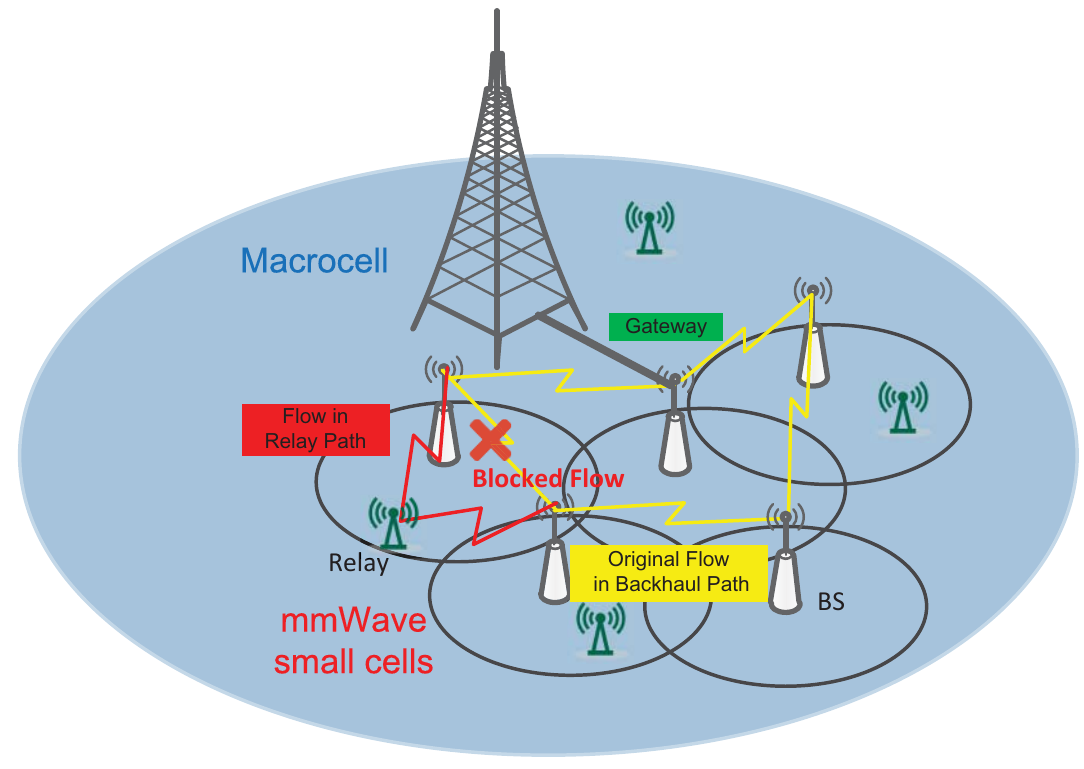}
  \caption{The small cells densely deployed underlying the macrocell.} \label{fig:smallcells}
  \end{center}
\end{figure}

We consider a scenario where small cells are densely deployed underlying the homogeneous macrocell. As shown in Figure \ref{fig:smallcells}, the network includes $N_B$ BSs and $N_R$ relays. There are one or more BSs connected to the backbone network via the macrocell, which is(are) called gateway(s) \cite{green}. A backhaul network controller (BNC) resides on one of the gateways. BNC synchronizes and coordinates the data transmission in the backhaul network \cite{survey}. It can obtain the QoS requirement of each flow and the location of each BS or relay. The BSs are connected through backhaul links in mmWave band to form a mesh network. When there is a traffic demand between two BSs, we say there is a flow between them. Each BS or relay is equipped with an electronically steerable directional antenna so that directional transmissions can be performed between the transmitters and receivers. When a flow is blocked, it can be forwarded through the surrounding relay nodes. For simplicity but without loss of generality, we just consider two-hop relay paths in mmWave band. In order to achieve a high transmission rate, in this paper, we assume that line of sight (LOS) transmissions can be achieved between the optional relays and the sources (or the destinations) of the blocked flows. Of course, the original flows (i.e. the unblocked flows) also perform LOS transmissions. Each node (BS or relay) is assumed to be half-duplex; so the flows sharing the common node can't be transmitted simultaneously.

\subsection{MAC Frame Structure}

\begin{figure}[bp]
  \begin{center}
  \includegraphics[width=7cm]{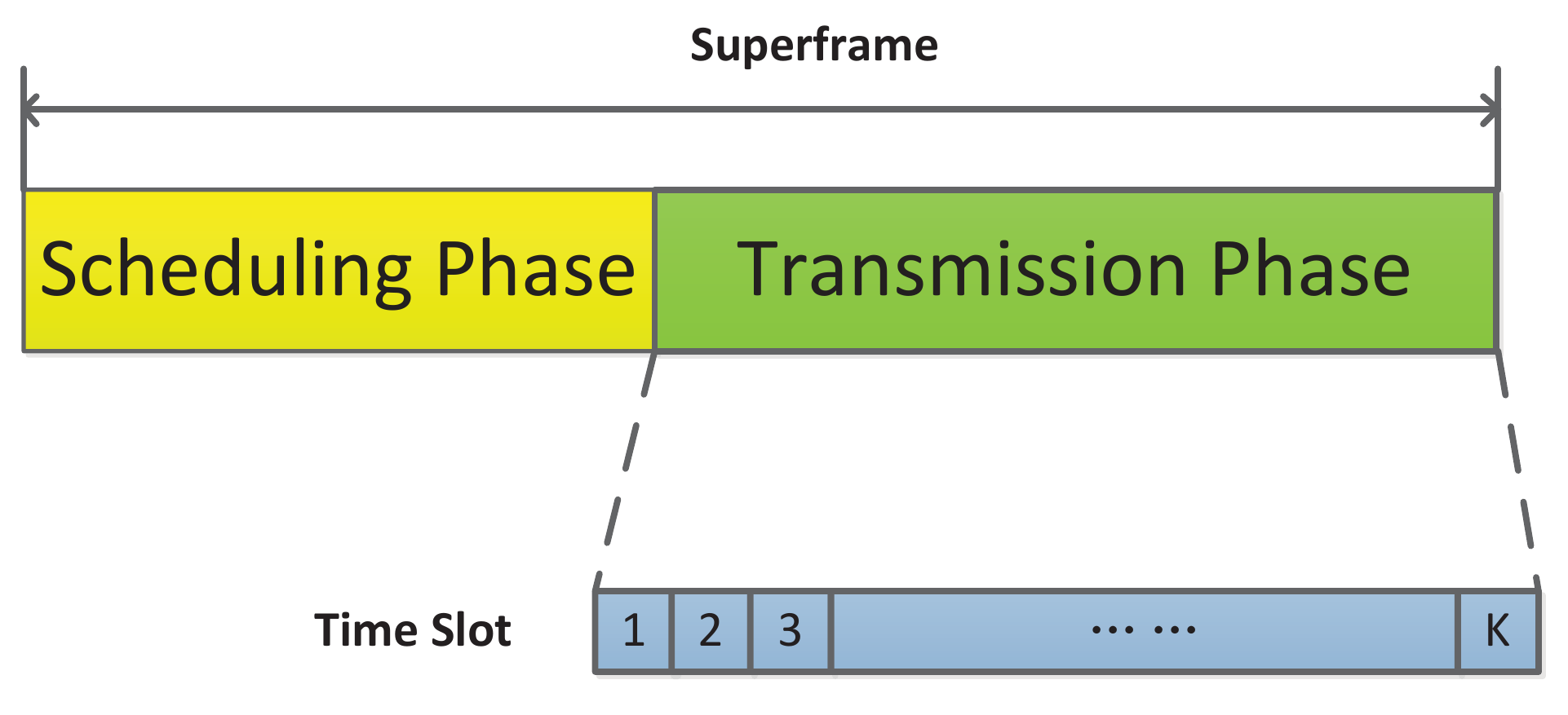}
  \caption{The structure of one superframe.} \label{fig:superframe}
  \end{center}
\end{figure}

In our algorithm, time is divided into a series of superframes \cite{survey}. As shown in Figure \ref{fig:superframe}, each superframe consists of two phases: scheduling phase and transmission phase \cite{frame}. In the scheduling phase, BNC receives the transmission request of each flow, selects relay(s) for the blocked flow(s) and makes the scheduling decision. Then, it broadcasts the scheduling decision to the whole network. In the transmission phase, time is further divided into $K$ equal time slots (TS). In every TS, some flows can be transmitted concurrently (either through a relay path or through a backhaul path) according to the scheduling decision.

\subsection{Received Power}
In this paper, we use a popular LOS path loss model for mmWave as described in \cite{pathloss}. The received power from the source node $s_f$ to the destination node $d_f$ of link $f$ can be expressed as
\begin{equation}
P_r\left(s_f,d_f\right)=kP_tG_t(s_f,d_f)G_r(s_f,d_f)d_{s_fd_f}^{-n}.\label{equation: received power}
\end{equation}
$k$ is a factor that is proportional to $\left(\frac{\lambda}{4\pi}\right)^{\tiny{2}}$, where $\lambda$ denotes the wave length; $P_t$ represents the transmission power of the transmitter; $G_t\left(s_f,d_f\right)$ represents the transmitted antenna gain in the direction of from $s_f$ to $d_f$ and $G_r\left(s_f,d_f\right)$ represents the received antenna gain in the direction of from $s_f$ to $d_f$, respectively; $d_{s_fd_f}$ denotes the distance between $s_f$ and $d_f$ and $n$ is the path loss exponent \cite{Qiao2}.

Similarly, the received interference from the source node $s_l$ of link $l$ to the destination node $d_f$ of link $f$ can be expressed as
\begin{equation}
P_r\left(s_l,d_f\right)=\rho kP_tG_t(s_l,d_f)G_r(s_l,d_f)d_{s_ld_f}^{-n},
\end{equation}
where $\rho$ is the multi-user interference (MUI) factor between different links, which is related to
the cross correlation of signals from different links \cite{zhuyun}.

\subsection{Data Rate}
With the reduction of multipath effect for directional mmWave links, the data rate of link \emph{f} can be estimated according to the Shannon's channel capacity \cite{green}.
\begin{equation}
R_f=\eta W{\rm{log}}_2(1+\frac{P_r(s_f,d_f)}{N_0W+\sum\limits_{l\not=f}P_r(s_l,d_f)}),\label{equation:rate}
\end{equation}
where $\eta$ is the factor that describes the efficiency of the transceiver design, which is in the range of $(0,1)$. $W$ is the channel bandwidth, and $N_0$ is the onesided power spectral density of white Gaussian noise \cite{Qiao2}. $l$ represents the link that is transmitted simultaneously with $f$. In fact, only when the scheduling decision is determined, or in other words, which links are scheduled at the same time with link $f$ is determined, the actual rate of one link can be determined.
\section{Problem Formulation}\label{S4}
In this section, we formulate the optimal scheduling problem when relay paths and backhaul paths coexist into an MINLP.

It's assumed that there are $F$ flows in the network and each flow \emph{f} has its own QoS requirement $q_f$. If a flow is blocked, it will select a relay path to forward data. The original flow is still transmitted through the backhaul path.

For flow \emph{f}, the maximum number of hops of the selected path is denoted as $H_{maxf}$. If it chooses the relay path, $H_{maxf}=2 $; if it is still transmitted in the backhaul path, $H_{maxf}=1$.
We use a binary variable $a_{fh}^i$ to indicate whether the \emph{h}th hop of the relay path for flow \emph{f} is scheduled in the \emph{i}th slot (\emph{i} = 1, 2, ... \emph{K}). If it is, $a_{fh}^i=1$; otherwise $a_{fh}^i=0$. The source and destination of the \emph{h}th hop of relay path for flow \emph{f} are denoted by $s_{fh}$ and $d_{fh}$, respectively. Similarly, $b_f^i$ indicates whether the backhaul path for flow \emph{f} is scheduled in the \emph{i}th slot. $s_f$ and $d_f$ denote the source and destination of the backhaul path for flow \emph{f}. Besides, binary variable $I(s_{fh},d_{fh},s_{lp},d_{lp})= 1$ means the \emph{h}th hop of the relay path for flow \emph{f} and the \emph{p}th hop of relay path for flow \emph{l} are adjacent (i.e. they share the common node); $I(s_f,d_f,s_l,d_l)= 1$ means the backhaul path for flow \emph{f} and flow \emph{l} are adjacent; $I(s_{fh},d_{fh},s_l,d_l)= 1$ means the \emph{h}th hop of the relay path for flow \emph{f} and the backhaul path for flow \emph{l} are adjacent.

In this paper, we aim at maximizing the number of flows satisfying their QoS requirements, i.e., the number of completed flows. This is because many applications in mmWave band require multi-Gbps throughput to guarantee transmission quality. As a result, the QoS requirements of flows should be taken into account. However, as the number of slots in the transmission phase is limited, if we blindly aim at increasing the total network throughput, the limited slot resources are always allocated to the flows with high transmission rates, so the flows with low transmission rates will hardly be scheduled, which is unfair. Therefore, aiming at maximizing the number of completed flows can ensure  both backhauling capacity and fairness.

For a blocked flow transmitted in relay path, only when the QoS requirement $q_f$ is achieved in both two hops, it can be called a completed flow. This can be expressed as $T_{f1} \ge q_f \And T_{f2} \ge q_f$, where $T_{f1}$ represents the actual throughput of the first hop of the relay path for flow $f$, and $T_{f2}$ represents the actual throughput of the second hop of the relay path for flow $f$.
For an original flow transmitted in backhaul path, when the QoS requirement $q_f$ is achieved in one hop, it is called a completed flow. This can be expressed as $T_{fb} \ge q_f$, where $T_{fb}$ represents the actual throughput of the backhaul path for flow $f$.
Specifically, the throughput of the link that is currently being scheduled in the superframe for flow $f$ can be expressed as
\begin{equation}
T_{f}=\frac{\sum\limits_{i=1}^KR_f^i\triangle t}{T_s+K\triangle t}.\label{equation:throughput}
\end{equation}
Here, $T_s$ is the time of scheduling phase and $\triangle t$ is the time of one slot. $R_f^i$ denotes the actual rate of flow \emph{f} in the \emph{i}th slot. The interference from other flows is considered. We use a scheduling vector $\textbf{c}^i$ to indicate which flow(s) is(are) scheduled in the \emph{i}th slot. In the vector, if the element $c_f^i= 1$, it means flow \emph{f} is scheduled in this slot; if $c_f^i= 0$, it means flow \emph{f} isn't scheduled. According to (\ref{equation:rate}), $R_f^i$ can be calculated as (\ref{equation:Rft}).
\begin{equation}
R_f^i=\eta W\mbox{${\rm{log}}_2$}(1+\frac{c_f^iP_r(s_f,d_f)}{N_0W+\sum\limits_{l\not=f}c_l^iP_r(s_l,d_f)})\label{equation:Rft}
\end{equation}

For convenience, we use a binary variable $I_f$ to indicate whether flow \emph{f} is completed. $I_f= 1$ indicates flow \emph{f} is completed; $I_f= 0$ indicates it isn't completed. Therefore, the optimal scheduling problem (P$1$) when the relay paths and backhaul paths coexist can be formulated as follows.
\begin{equation}
\max\sum\limits_{f=1}^FI_f   \label{goal}
\end{equation}
For a blocked flow transmitted in relay path,
\begin{equation}
I_f=
\begin{cases}
1, &\mbox{$T_{f1} \ge q_f \And T_{f2} \ge q_f$;}\\
0, &\mbox{otherwise.}
\end{cases}
\end{equation}
For an original flow transmitted in backhaul path,
\begin{equation}
I_f=
\begin{cases}
1, &\mbox{$T_{fb} \ge q_f$;}\\
0, &\mbox{otherwise.}
\end{cases}
\end{equation}

Now let's analyze the constraints. First, due to the half-duplex nature of the node, adjacent links can't be transmitted simultaneously. Here, three cases are included: 1) when both two adjacent flows are blocked and transmitted in the relay paths, the constraint can be expressed as
\begin{equation}
\begin{aligned}
a_{fh}^i+a_{lp}^i\le1,  \mbox{  if } I(s_{fh},d_{fh},s_{lp},d_{lp})= 1;
\mbox{  }\forall f,l,h,p,i; \label{equation: adjacent1}
\end{aligned}
\end{equation}
2) when both two adjacent flows are not blocked and transmitted in the backhaul paths, the constraint can be expressed as
\begin{equation}
b_f^i+b_l^i\le1,  \mbox{  if } I(s_f,d_f,s_l,d_l)= 1;
\mbox{  }\forall f,l,i; \label{equation: adjacent2}
\end{equation}
3) when one flow is transmitted in the relay path and its adjacent flow is transmitted in the backhaul path, the constraint can be expressed as
\begin{equation}
a_{fh}^i+b_l^i\le1,  \mbox{  if } I(s_{fh},d_{fh},s_l,d_l)= 1;
\mbox{  }\forall f,l,h,i. \label{equation: adjacent3}
\end{equation}

Second, if flow $f$ selects the relay path, due to the inherent order of transmission, different hops in the same path can't be concurrently scheduled, which can be expressed as
\begin{equation}
\sum\limits_{h=1}^{H_{maxf}}a_{fh}^i\le 1;\mbox{  }\forall f,i. \label{equation: samepath}
\end{equation}

Third, in the relay path, the \emph{h}th hop should be scheduled ahead of the $\left(h+1\right)$th hop due to the inherent transmission order, which can be expressed as
\begin{equation}
\begin{aligned}
\sum\limits_{i=1}^{T^*}a_{fh}^i\ge \sum\limits_{i=1}^{T^*}a_{f\left(h+1\right)}^i,\mbox{  if }H_{maxf}>1;\\
\forall h = 1 \sim \left(H_{maxf}-1\right), T^*=1 \sim K. \label{equation:h ahead h plus 1}
\end{aligned}
\end{equation}
Note that constraint (\ref{equation:h ahead h plus 1}) is a group of constraints, since $T^*$ varies from
1 to $K$. Besides, $h$ varies from 1 to $H_{maxf}-1$, which ensures that each prior hop is scheduled ahead of the hop behind.

Finally, for one flow, it can only select one path at most. In other words, in one slot, it is transmitted either in the  backhaul path or in one hop of the relay path. If the flow is blocked but it doesn't select a relay node, it can't be transmitted at all, which can be expressed as
\begin{equation}
a_{fh}^i+b_f^i
\begin{cases}
\le1, &\mbox{  if $q_f>0\And h<H_{maxf}$;}\\
=0, &\mbox{  otherwise;}
\end{cases}
\mbox{  }\forall f,h,i.\label{equation: backhaul and relay}
\end{equation}

This is a mixed integer nonlinear programming problem (MINLP) and is NP-hard. It's complex and is difficult to be solved in polynomial time. Therefore, we should propose an efficient and pratical algorithm to solve it.
\section{The Proposed RAQS Scheme} \label{S5}
In this section, we describe the proposed relay-assisted and QoS aware scheduling scheme (RAQS). It mainly includes two parts. The first part is a relay selection algorithm for the blocked flows, and the second part is a heuristic transmission scheduling algorithm when the relay paths and backhaul paths coexist. The concurrent transmissions are fully exploited and the QoS requirements of flows are especially considered in both two parts. In particular, it's assumed that during the transmission, the blockage is always here. Note that in this paper, when the QoS requirement of one flow is achieved, the flow is called a completed flow.
\subsection{Relay Selection Algorithm}
When a flow is blocked, we need to select a relay for it to forward data. There are multiple relay nodes in the scenario. However, it's not true that any relay is beneficial to achieve the QoS requirements of more flows. On one hand, if the rate of the relay link is too low, even if the flow can be transmitted throughout the transmission phase, it's not necessarily able to achieve its QoS requirement in the limited superframe time. On the other hand, if multiple blocked flows select the same relay, it may lead to more node contentions because every node is half-duplex. This is not conducive to concurrent transmissions and thus reduces the transmission efficiency. As a result, it is also not beneficial to satisfy more flows' QoS requirements.

\begin{figure}[tbp]
  \begin{center}
  \includegraphics[width=7cm]{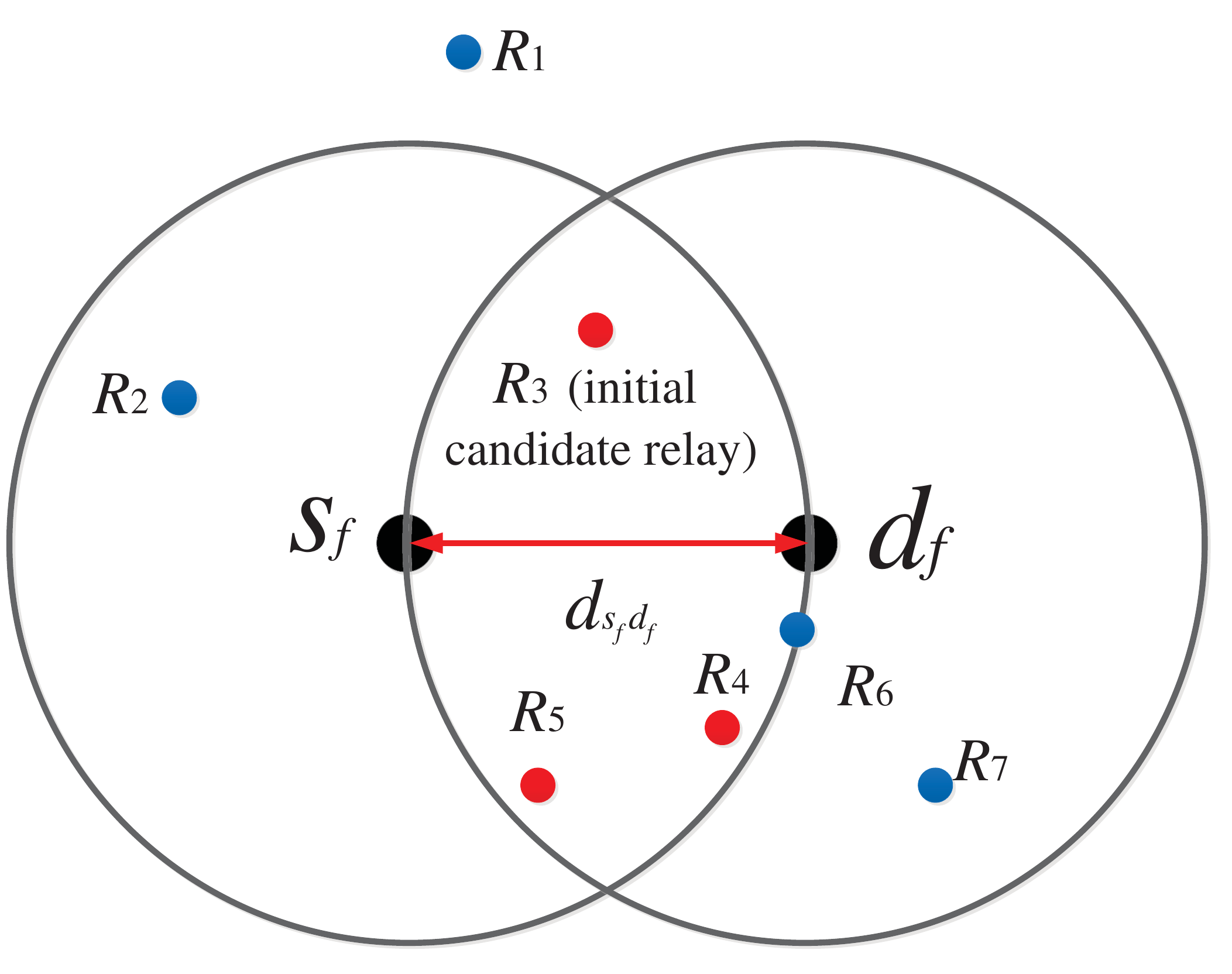}
  \caption{The selection of initial candidate relays for the blocked flow from $s_f$ to $d_f$.} \label{fig:relay}
  \end{center}
\end{figure}

To guarantee a high rate, according to (\ref{equation: received power}) and (\ref{equation:rate}), the selected relay(s) can't be too far from the source and the destination of the blocked flow. As shown in Fig. \ref{fig:relay}, if the flow $f$ between $s_f$ and $d_f$ is blocked, we draw two circles with $s_f$ and $d_f$ as the centers, respectively. The radiuses of both circles are equal to the distance $d_{s_fd_f}$ between $s_f$ and $d_f$. The relay nodes that fall within the overlap of the two circles ($R_3, R_4$ and $R_5$, does not include the borders) become the initial candidate relay set for flow $f$, which is denoted as $\textbf{Can1}(f)$.

In order to further guarantee the transmission rate of flow $f$, we can then select relay(s) from $\textbf{Can1}(f)$ according to the time that it takes for the relay path to transmit a certain amount of data. Only the relays whose used time meets some condition can be selected. The condition can be described as
\begin{equation}
\frac{\frac{1}{R_b}}{\frac{1}{R_1}+\frac{1}{R_2}}>\beta,\label{eqa: beta}
\end{equation}
where $R_b$ denotes the rate of the backhaul path when the flow is not blocked; $R_1$ and $R_2$ denote the rates of the first and the second hop of the relay path, respectively. All rates here are calculated without interference, because we have not made the scheduling decision yet and the interference can't be determined. $\beta$ is called the relay selection parameter, which can be adjusted according to the actual situation. When a certain amount of data $D$ is transmitted, for the backhaul path, the time it takes can be expressed as $\frac{D}{R_b}$; for the two-hop relay path, the time it takes can be expressed as $\frac{D}{R_1}+\frac{D}{R_2}$. So the formula on the left side of the greater-than sign represents the time ratio between the backhaul path and relay path. To simplify the subsequent description, we call it \emph{TR} (time ratio). The relay set for flow $f$ selected in this way is denoted as $\textbf{Can2}(f)$. If there are more than one relays in $\textbf{Can2}(f)$, we then choose the relay with the maximum \emph{TR} in $\textbf{Can2}(f)$ and denote it by $\textbf{Can3}(f)$.

It's worth noting that if in \textbf{Can3}, different flows select the same relay, because of the half-duplex nature, there may be more node contentions, which is harmful to concurrent transmissions and achieve the QoS requirements of more flows. Therefore, we only assign the repeated relay to the flow that needs it most.

\begin{algorithm}[htbp]
\DontPrintSemicolon
\caption{Eliminating the Repeated Relay} \label{alg:Eliminating the repeated relay}
\textbf{Input:} The existing candidate relay set array \textbf{Can2}; The existing selected path array \textbf{P};\\
~~~~~~~~~The two flows $f_1$, $f_2$ that select the same relay $r$; \\
\textbf{Output:} The new selected path array $\textbf{P}$ and new \textbf{Can2};\\
set $n_1=$ the length of $\textbf{Can2}(f_1)$, $n_2=$ the length of $\textbf{Can2}(f_2)$;\\
\If{$n_1==1\And n_2==1$}
{
$\textbf{P}\left(f_i\right)=r$, $f_i$ is the flow with a higher $TR$; $f_j$ removes $r$ from $\textbf{Can2}(f_j)$; $\textbf{P}\left(f_j\right)=0$;\\
}
\ElseIf{$n_1==1\And n_2>1$}
{
$\textbf{P}\left(f_1\right)=r$; $f_2$ removes $r$ from $\textbf{Can2}(f_2)$;\\
\If{$\textbf{\rm{Can2}}(f_2)\not=\emptyset$}
{
 $\textbf{P}\left(f_2\right)=$ the suboptimal relay $r^{'}$;\\
\If{$r^{'}$ \rm{has} \rm{been} \rm{assigned} \rm{to} $f_3$}
{
$f_1=f_3$, iterate Algorithm 1;\\
}
}
\Else
{
$\textbf{P}\left(f_2\right)=0$;\\
}
}
\ElseIf{$n_1>1\And n_2==1$}
{
$\textbf{P}\left(f_2\right)=r$, $f_1$ removes $r$ from $\textbf{Can2}(f_1)$; \\
\If{$\textbf{\rm{Can2}}(f_1)\not=\emptyset$}
{
$\textbf{P}\left(f_1\right)=$ the suboptimal relay $r^{'}$;\\
\If{$r^{'}$ \rm{has} \rm{been} \rm{assigned} \rm{to} $f_3$}
{
$f_2=f_3$, iterate Algorithm 1;\\
}
}
\Else
{
$\textbf{P}\left(f_1\right)=0$;\\
}
}
\Else
{
$\textbf{P}\left(f_i\right)=r$, $f_i$ is the flow with a higher $TR$; the other $f_j$ removes $r$ from $\textbf{Can2}(f_j)$; \\
\If{$\textbf{\rm{Can2}}(f_j)\not=\emptyset$}
{
$\textbf{P}\left(f_j\right)=$ the suboptimal relay $r^{'}$;\\
\If{$r^{'}$ \rm{has} \rm{been} \rm{assigned} \rm{to} $f_3$}
{
$f_i=f_3$, iterate Algorithm 1;\\
}
}
\Else{
$\textbf{P}\left(f_j\right)=0$;\\
}
}
\end{algorithm}

The algorithm that eliminating the repeated relay is shown in Algorithm \ref{alg:Eliminating the repeated relay}. We allocate the relay according to the number of relays in \textbf{Can2}, because the relays in \textbf{Can2} are the relays with high rates. For a flow $f$, if there is only one relay in $\textbf{Can2}(f)$, we think it needs the relay most. Otherwise, a higher \emph{TR} means the flow needs the relay more. In addition, we denote the selected path as \textbf{P}. If a blocked flow \emph{f} picks out a relay, $\textbf{P}(f)$ is set to be the selected relay's ID (1, 2...$N_R$). Otherwise $\textbf{P}(f)=0$. The initial \textbf{P} is equal to \textbf{Can3}. Specifically, the following three cases are included. 1) If the two flows that select the same relay both have only one candidate relay in \textbf{Can2}, we assign the relay to the flow with a higher \emph{TR}, denoted as $f_i$. Therefore, the other flow $f_j$ can't be transmitted, which is described in line 6. 2) If one of the two flows has only one candidate relay in \textbf{Can2}, and the other has more than one relay in \textbf{Can2}, we assign the relay to the former. The latter removes the repeated relay from \textbf{Can2} and selects the suboptimal relay (i.e. the relay with the second highest \emph{TR} in \textbf{Can2}, as shown line 7-22. 3) If both of the two flows have more than one candidate relay in \textbf{Can2}, we also assign the relay to the flow with a higher \emph{TR}. The other removes the repeated relay from \textbf{Can2} and selects the suboptimal relay, as shown in line 23-30. If the suboptimal relay has been assigned to other flow(s), we iteratively execute the above three rules until an unused relay is selected for the flow or its \textbf{Can2} becomes empty. \textbf{Can2} becomes empty means the flow doesn't have a path to transmit.

\subsection{The Proposed Transmission Scheduling Algorithm}
After selecting proper relay for each blocked flow as indicated by constraint (\ref{equation: backhaul and relay}), we propose a heuristic scheduling algorithm to solve the joint scheduling problem of relay paths and backhaul paths. In order to fully exploit concurrent transmissions and let more flows achieve their QoS requirements, the concept of contention graph in \cite{contention graph} is still used. The contention graph could reflect the global information of the contentions residing in the network \cite{zhuyun}. Besides, the difference between two-hop relay path and one-hop backhaul path is fully considered.

\begin{figure}[bp]
  \begin{center}
  \includegraphics[width=4cm]{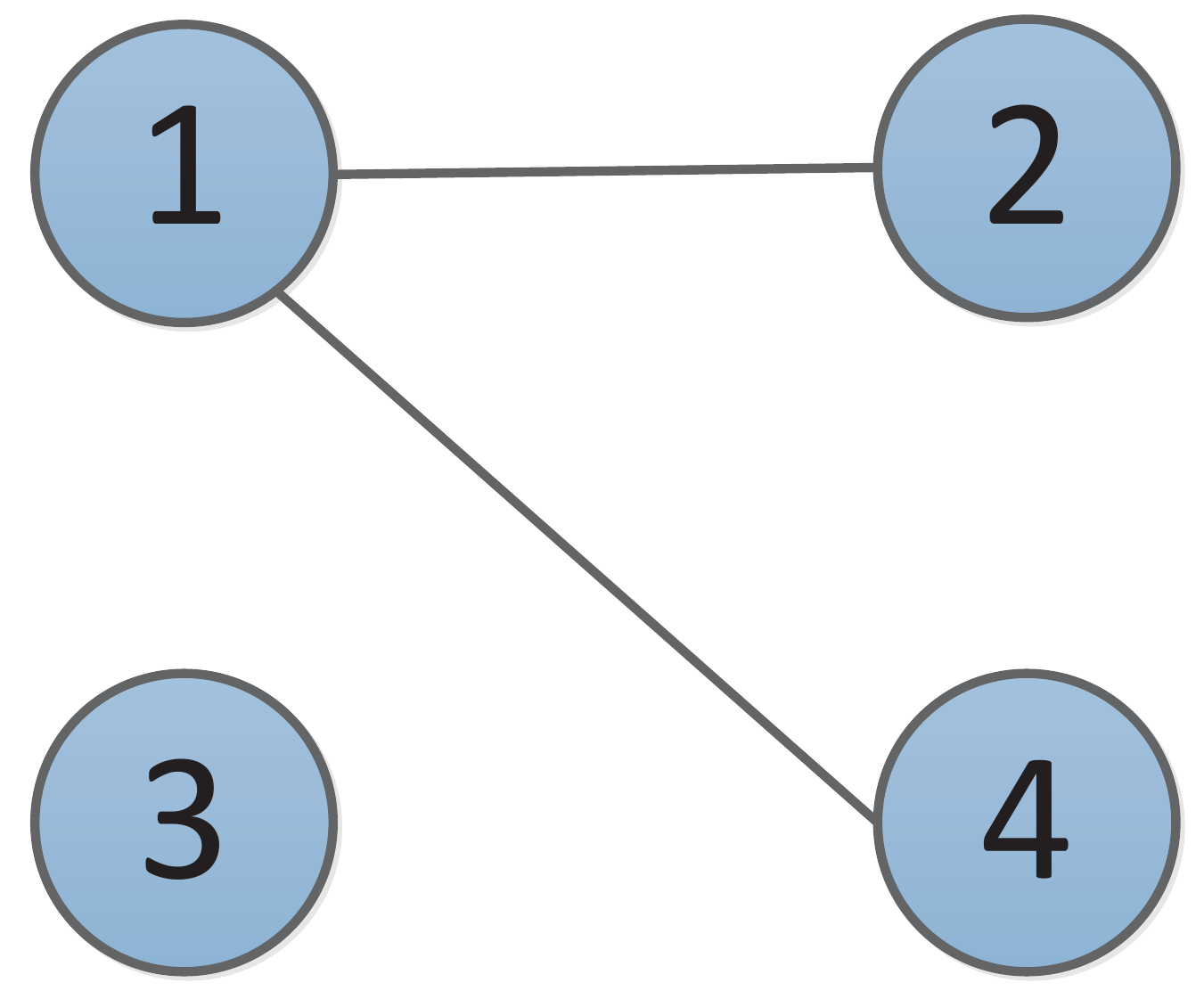}
  \caption{The contention graph with contention between link 1 and link 2 and contention between link 1 and link 4.} \label{fig:contentionGraph}
  \end{center}
\end{figure}

In the contention graph, each vertex represents one link (relay link or backhaul link). If two links share the common node, or the interference that one link has on another is bigger than a threshold $\sigma$, as shown in (\ref{equation:contention}), we say there is a contention between them and then add one edge between the two vertices. Links that have common node cannot be scheduled concurrently due to the constraints (\ref{equation: adjacent1}), (\ref{equation: adjacent2}), and (\ref{equation: adjacent3}).
For example, as shown in Figure \ref{fig:contentionGraph}, there is a contention between link 1 and link 2, which indicates link 1 and link 2 share the common node or the interference between them is severe. The links with contention can't be scheduled simultaneously. In contrast, there is no contention between link 1 and link 3. So they can be scheduled in the same slot. The edge number of one vertex is called the degree of the link. For instance, the degree of link 1 is 2, and the degree of link 3 is 0.

\begin{equation}
\max \{P_r(s_l,d_f),P_r(s_f,d_l)\}>\sigma\label{equation:contention}
\end{equation}

The transmission scheduling algorithm is shown in Algorithm \ref{alg:The scheduling algorithm}. Firstly, in line 3, BNC receives the transmission request of each flow with their QoS requirements $q_f$, and then it calculates the total number of slots $\xi_f$ that flow $f$ spends achieving its QoS requirement when transmitted in the selected path. If the flow is transmitted in the relay path, the total number of slots $\xi_f$ is equal to the sum of the number of slots spent in each hop, which can be expressed as $\xi_f = \xi_1 + \xi_2$. $\xi_1$ represents the number of slots spent in the first hop of the relay path, and $\xi_2$ represents the number of slots spent in the second hop of the relay path. If the flow is transmitted in the backhaul path, $\xi_f$ is equal to the number of slots spent in the only one hop, denoted as $\xi_f = \xi_b$. $\xi_b$ denotes the number of slots spent in the one-hop backhaul path. Specifically, the number of slots $\xi$ spent in the current hop can be calculated as (\ref{equation: slot}). $R_f$ is the rate of the current hop with no interference from other links. The numerator represents the total number of bits that need to be transmitted in one superframe. The denominator represents the number of bits that flow $f$ can transmit in one slot in the current hop.
\begin{equation}
\xi = \frac{q_f*\left(T_s+K\triangle t\right)}{R_f*\triangle t}\label{equation: slot}
\end{equation}

The flows that spend too many slots will be removed from the scheduling set, as shown in line 4. This is based on the knowledge that the number of slots in the transmission phase is limited. If the flow has been scheduled throughout the transmission phase, but it still can't achieve it's QoS requirement, the slots are wasted. Next, in line 6, we initialize the unscheduled headmost hop $F_f$ and the maximum number of hops $H_{maxf}$ for each flow $f$. $F_f$ is set to 1 at the beginning, which ensures the first hop of flows transmitted in the relay path is scheduled first indicated by constraints (\ref{equation: samepath}) and (\ref{equation:h ahead h plus 1}).
In line 7, We also initialize a $F\times K$ scheduling matrix $\textbf{C}=[\textbf{c}^1,\textbf{c}^2,... ...\textbf{c}^K]$, denoted the scheduling decision in $K$ slots.

\begin{algorithm}[htbp]
\DontPrintSemicolon
\caption{The Transmission Scheduling Algorithm} \label{alg:The scheduling algorithm}
\textbf{Input:}The final selected path array \textbf{P};\\
\textbf{Output:}The scheduling matrix $\textbf{C}$;\\
BNC receives the transmission request of each flow with their QoS requirements $q_f$ and calculates $\xi_f$ of each flow;\\
remove $\mathbb{D}=\{f|\xi_f>K\}$;
$F$ = the number of remaining flows;\\
\textbf{Initialization:} $F_f$ = 1 and $H_{max_f}$ for each flow;
~$\textbf{C}_{\scriptsize{F\times K}} =\textbf{0}$; \\
\For {\rm{slot} $i\left(1\le i\le K\right)$}
{
\If {$i=1$ \rm{or} \rm{one} \rm{hop} \rm{of} \rm{some} \rm{flow} \rm{is} \rm{newly} \rm{completed}}
{
generate \textbf{G} of all $F$ flows in the current hop and remove invalid flows from \textbf{G};\\
\While{$\bf{G}\not=\emptyset $}
{
$\mathbb{L}$= the set of remaining flows in \textbf{G}; \\
obtain $F_f$ for the flows in $\mathbb{L}$;\\
$\mathbb{T}_{\mbox{\scriptsize{wo}}} = \{l|F_l==2\}$;\\
\If{$\left|\mathbb{T}_{\mbox{\scriptsize{wo}}}\right|>1$}
{
$\mathbb{M} = \{t|\min \limits_{t\in\mathbb{T}_{\mbox{\tiny{wo}}}}\textbf{degree}\left(t\right)\}$;\\
\If{$\left|\mathbb{M}\right|>1$}{
$f=\min \limits_{f\in\mathbb{M}}\xi$, $\xi$ is the number of slots spent in the current $F_f$th hop;\\
}
\Else
{
select $f\in \mathbb{M}$;\\
}
}
\Else
{
select $f\in \mathbb{T}_{\mbox{\scriptsize{wo}}}$;\\
}
$c_f^i=1$;\\
remove $f$ and its neighbors from \textbf{G};
}
}
\Else{
$\textbf{c}^{i}=\textbf{c}^{i-1}$;\\
}
\If {\rm{any} $T_{f}>q_f$}
{
\If{$F_f=H_{max_f}$}
{
 $c_f^i=-1$;\\
 }
\Else
{
$F_f=F_f+1$;\\
}
}
}
\end{algorithm}

We make the scheduling decision slot by slot. If it is the first slot or some flows newly achieve their QoS requirements in the current hop, we use the method in \cite{contention graph} to generate the contention graph \textbf{G} of all $F$ flows in the current hop. The flows that have been completed and that are ongoing don't need to be judged again. To avoid contention, the neighbor(s) of ongoing flow(s) in contention graph shouldn't be scheduled. These three kinds of flows are called invalid flows and we remove them from \textbf{G}, as shown in line 8. Then, in line 9-22, based on \textbf{G}, we make the scheduling decision. While \textbf{G} is not empty, we prefer to select the flow whose current hop $F_f$ equals 2, as shown in line 12. This is because that it means the first hop has been finished, if we don't schedule the second hop, the slots used in the first hop are wasted. However, if there are multiple flows whose current hops equal 2, the flow that has the minimal degree is preferred, which is shown in line 13-14. Smaller degree means there is less node contention or smaller interference between this link and other links, which is beneficial to concurrent transmissions and satisfy more flows' QoS requirements, which is the objective function in (\ref{goal}). If there are still multiple flows that have the same minimal degree, we select the flow that spends the least number of slots in the current hop, as shown in line 15-16. The faster one flow achieves its QoS requirement, the more slots can be saved to let other flows be transmitted, which is beneficial to satisfy the QoS requirements of more flows. The process of selecting a transmission flow is shown in line 12-21. Then the newly selected flow and its neighbor(s) are also removed from the contention graph, as shown in 22. We repeat these steps until the contention graph becomes empty. In this way, we can pick out all the flows that are scheduled in slot $i$, denoted as $\textbf{c}^i$. If no flow is completed, as shown in line 24, we still use the scheduling decision of the last slot. At the end of each slot, as shown in line 25-29, we should check whether there are some flows achieve their QoS requirements in the current hop. If the flow $f$ has achieved $q_f$ and the current hop is the maximum hop, it is completed and will never be scheduled later, which is denoted as $c_f^i=-1$. It helps us to save slots and satisfy more flows' QoS requirements. If it isn't the maximum hop, increase the hop $F_f$ by 1. As long as one hop of some flow is transmitted completely, the contention graph in the current hop needs to be regenerated and the number of slots spent in the current hop needs to be recalculated. We iteratively make decisions with the method described above, until all slots are completed.

To estimate the algorithm complexity, we can observe the outer for loop has $K$ iterations. The inner while loop has $F$ iterations in the worst case, Thus, the scheduling algorithm has the complexity of $O(FK)$, which can be implemented in practice.

\begin{table}[htbp]
\caption{Simulation parameters.} \label{table:parameter setting}
\centering  
\begin{tabular}{lccc}  
\hline
\textbf{Parameter} &\textbf{Symbol}&\textbf{Value}\\ \hline  
Transmission power &$P_t$ &1000mW\\
Path loss exponent  &$n$ &2\\
MUI factor &$\rho$ &1\\
Transceiver efficiency factor &$\eta$ &0.5\\
System bandwidth &$W$ &1200MHz\\
Background noise &$N_0$ &-134dbm$/$MHz\\     
Slot time &$\triangle t$ &18us\\
Scheduling phase time &$T_s$ &850us\\
Number of slots in transmission phase &$K$ &3000\\
Half-power beamwidth &$\theta_{\mbox\scriptsize{-3dB}}$ &$30^\circ$\\
 \hline
\end{tabular}
\end{table}

\section{Performance Analysis}\label{S5-b}

In this section, we analyze the impact of interference threshold choice on the performance of our scheme.
To fully reap the benefits of concurrent transmissions, the sum of transmission rates of links scheduled for transmission in the same time slot
should be maximized. This sum can also be regarded as the throughput in one time slot, and has a big impact on the system performance.
We denote the set of concurrent links scheduled in the $i$th slot as $V_i$. For one link $f\in V_i$, we can obtain its transmission rate as

\begin{equation}
{R_f} = \eta W{\rm{lo}}{{\rm{g}}_2}\left( {1 + \frac{{{P_r}({s_f},{d_f})}}{{{N_0}W + \sum\limits_{l\not  = f,l \in {V_i}} {{P_r}} ({s_l},{d_f})}}} \right).\label{equation:rate}
\end{equation}
The sum of transmission rates of links scheduled in the $i$th slot can be obtained as
\begin{equation}
\sum\limits_{f \in {V_i}} {{R_f} = } \sum\limits_{f \in {V_i}} {\eta W{\rm{lo}}{{\rm{g}}_2}\left( {1 + \frac{{{P_r}({s_f},{d_f})}}{{{N_0}W + \sum\limits_{l\not  = f,l \in {V_i}} {{P_r}} ({s_l},{d_f})}}} \right)}.
\end{equation}
As stated before, concurrent links should have no contention. As indicated in (\ref{equation:contention}), the interference between concurrent links
is less than or equal to $\sigma$. Thus, the sum rate meets
\begin{equation}
\sum\limits_{f \in {V_i}} {{R_f} \ge } \sum\limits_{f \in {V_i}} {\eta W{\rm{lo}}{{\rm{g}}_2}\left( {1 + \frac{{{P_r}({s_f},{d_f})}}{{{N_0}W + (|{V_i}| - 1)\sigma }}} \right)}.\label{sum-rate}
\end{equation}
The right side of (\ref{sum-rate}) can be regarded as a lower bound of the sum rate. To maximize the sum rate, we can optimize the interference threshold $\sigma$
to maximize the lower bound, which can be expressed as
\begin{equation}
\begin{aligned}
&\sum\limits_{f \in {V_i}} {\eta W{\rm{lo}}{{\rm{g}}_2}\left( {1 + \frac{{{P_r}({s_f},{d_f})}}{{{N_0}W + (|{V_i}| - 1)\sigma }}} \right)}  \\&= \eta W{\rm{lo}}{{\rm{g}}_2}\prod\limits_{f \in {V_i}} {\left( {1 + \frac{{{P_r}({s_f},{d_f})}}{{{N_0}W + (|{V_i}| - 1)\sigma }}} \right)}.
\end{aligned}
\end{equation}
To maximize the lower bound, we should maximize $\prod\limits_{f \in {V_i}} {\left( {1 + \frac{{{P_r}({s_f},{d_f})}}{{{N_0}W + (|{V_i}| - 1)\sigma }}} \right)} $.
The number of concurrent links ${|{V_i}|}$ is determined by the threshold $\sigma$. When $\sigma$ increases, more links will have no contention between each other.
Thus, ${|{V_i}|}$ also increases, and the number of product terms increases. However, each product term will decrease. When $\sigma$ decreases, ${|{V_i}|}$ also decreases.
The number of product terms decreases, while each product term will increase. Therefore, both too large and too small $\sigma$ will decrease the sum rate. There should be
an optimized value of $\sigma$ that can maximize the sum rate, which is consistent with the performance evaluation results in Fig. \ref{fig:threshold1} and Fig. \ref{fig:threshold2}.

On the other hand, since the ${\rm{log}}_2(x)$ function is convex, we can obtain

\begin{equation}
\begin{aligned}
&\eta W\sum\limits_{f \in {V_i}} {{\rm{lo}}{{\rm{g}}_2}\left( {1 + \frac{{{P_r}({s_f},{d_f})}}{{{N_0}W + (|{V_i}| - 1)\sigma }}} \right)} \\& \le \eta W|{V_i}|{\rm{lo}}{{\rm{g}}_2}\left( {1 + \sum\limits_{f \in {V_i}} {\frac{{{P_r}({s_f},{d_f})}}{{{N_0}W|{V_i}| + |{V_i}|(|{V_i}| - 1)\sigma }}} } \right).
\end{aligned}
\end{equation}
The equal sign is taken when ${\frac{{{P_r}({s_f},{d_f})}}{{{N_0}W + (|{V_i}| - 1)\sigma }}}$ is equal for each link $f\in V_i$. When $\sigma$ and $|{V_i}|$
is fixed, more uniform ${{P_r}({s_f},{d_f})}$ can achieve higher sum rate and thus better network performance. With the transmission power fixed, more uniform link length
can achieve better performance. Therefore, the relays should be deployed to form uniform relay link length as the backhaul link to achieve better performance.
Thus, $\beta$ should be set to near 0.5 to achieve a better network performance, which is also indicated in Fig. \ref{fig:slv1} and Fig. \ref{fig:slv2}.

\section{Performance Evaluation} \label{S6}
\subsection{Simulation Setup}

In the simulations, as the algorithm performance is dependent on the locations of BSs and relays, we consider a scenario that 10 BSs are uniformly and randomly distributed in a $100m\times100m$ square area. The relay nodes obey space poisson distribution with parameter $\lambda = 30$. The number of flows is set to 10. The sources and destinations of 10 flows are randomly selected. The QoS requirement of each flow is uniformly distributed between 1Gbps and 3Gbps \cite{zhuyun}. The blocked flow(s) is(are) also randomly set and the frequency of mmWave is 60GHz. Both BSs and relays have the same transmission power $P_t$. Particularly, we use the realistic antenna model in \cite{antenna model1}. The gain of a directional antenna in units of dB can be expressed as follows.
\begin{equation}
G(\theta) =
\begin{cases}
G_0-3.01\times\left(\frac{2\theta}{\theta_{\mbox{\tiny{-3dB}}}}\right)^2, &\mbox{$0^{\circ}\le\theta\le\theta_{ml}/2$}\\
G_{sl}. &\mbox{$\theta_{ml}/2<\theta\le180^{\circ}$}
\end{cases}
\end{equation}
$\theta$ denotes an angle within the range $[0^\circ,180^\circ]$. $G_0$ is the maximum antenna gain and it can be expressed as $G_0=\mbox{10log}(1.6162/\mbox{sin}(\theta_{{\mbox{\scriptsize{-3dB}}}}/2))^2$. $\theta_{{\mbox{\scriptsize{-3dB}}}}$ denotes the angle of the half-power beamwidth. $\theta_{ml}$ denotes the main lobe width in units of degrees and it can be expressed as $\theta_{ml}=2.6\times\theta_{\mbox{\scriptsize{-3dB}}}$. The sidelobe gain $ G_{sl}=-0.4111\times\mbox{ln}(\theta_{\mbox{\scriptsize{-3dB}}})-10.579$ \cite{green}. To better simulate the real scenario, we choose other relevant parameters as shown in Table \ref {table:parameter setting}, most of which are the same as those in \cite{Qiao2}.

\subsection{Schemes for Comparison and Metrics for Evaluation}

In the simulations, we compare our RAQS algorithm with the following three schemes:

1) \emph{\textbf{MQIS}}: The maximum QoS aware independent set based scheduling algorithm. In the
algorithm, concurrent transmissions and the QoS aware priority are exploited to achieve more successfully scheduled flows and higher network throughput \cite{zhuyun}. To best of our knowledge, MQIS achieves the best performance in terms of the number of flows satisfying their QoS requirements and the system throughput. However, it doesn't provide a solution to the blockage problem.

2) \emph{\textbf{STDMA}}: The spatial-time division multiple access algorithm \cite{Qiao2}. In this algorithm, if scheduling one flow can increase the system throughput, we then decide to schedule it. Similarly, it still doesn't provide a solution to the blockage problem.

3) \emph{\textbf{Random relay}}: For the blocked flows, the random relay selection algorithm selects the final relay(s) without any special algorithm. It just selects the final relay(s) uniformly and randomly.

The two metrics, number of completed flows and system throughput, are used for evaluation. Only when a flow achieves its QoS requirement in all hops of the selected path, can it be called a completed flow.
The number of completed flows is the number of completed flows in the system until the end of simulation.
The system throughput represents the throughput of all flows in the network, which also includes the throughput of uncompleted flows.

Particularly, we simulate these two metrics under different number of blocked flows and different interference thresholds $\sigma$ in the contention graph. Besides, the impact of relay selection parameter $\beta$ in (\ref{eqa: beta}) is also simulated. The simulations are repeated 100 times to get the average results.

\subsection{Simulation Results}

\begin{figure}[htbp]
\vspace*{-2mm}
\begin{minipage}[t]{1\linewidth}
\centering
\includegraphics[width=0.85\columnwidth]{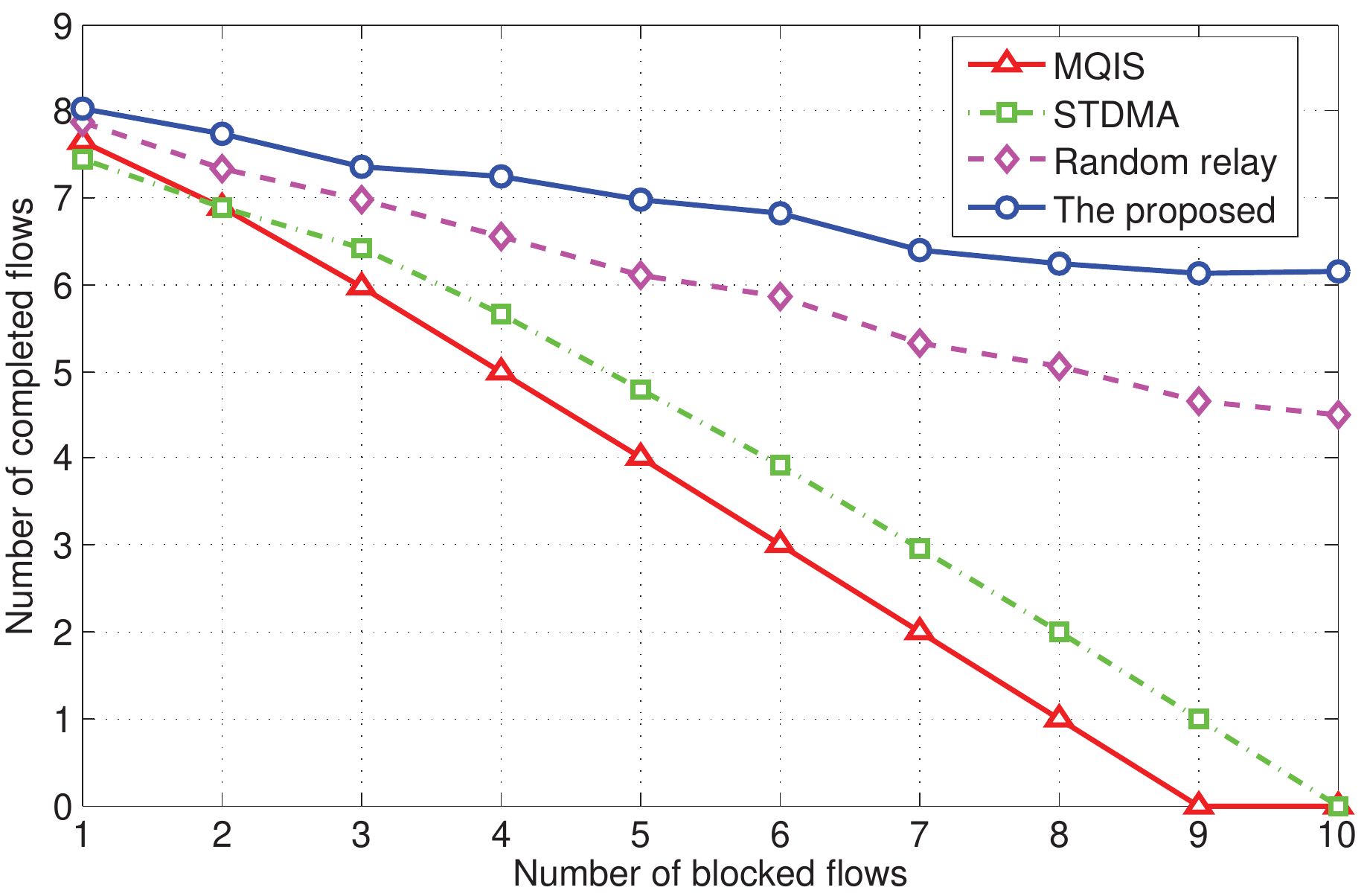}
\end{minipage}%
\vspace*{-3mm}
\caption{Number of completed flows
under different number of blocked flows.}
\label{fig:QoS_off}
\vspace*{0mm}
\end{figure}

\begin{figure}[htbp]
\vspace*{-2mm}
\begin{minipage}[t]{1\linewidth}
\centering
\includegraphics[width=0.85\columnwidth]{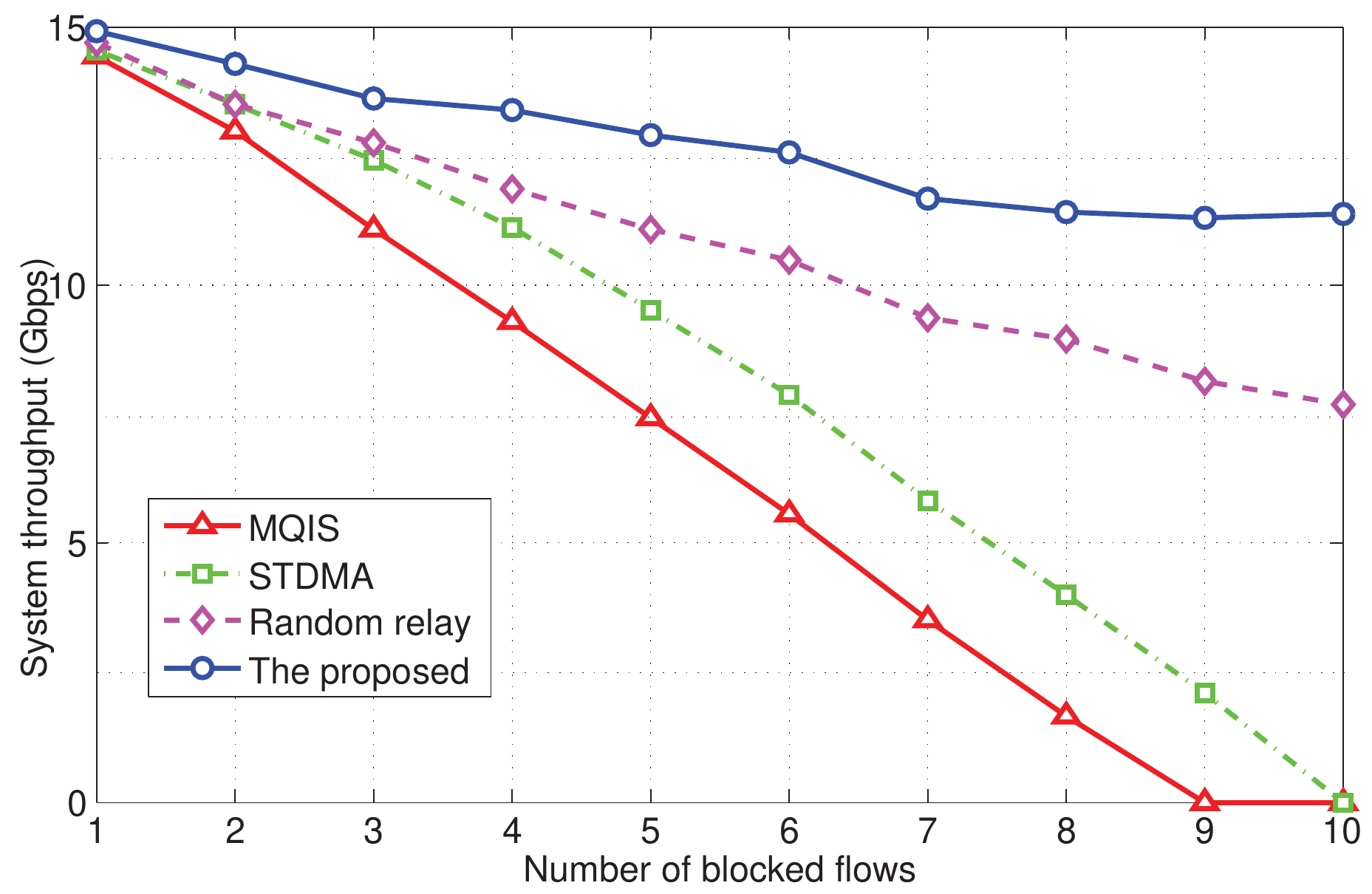}
\end{minipage}%
\vspace*{-3mm}
\caption{System throughput
under different number of blocked flows.}
\label{fig:thro_off}
\vspace*{-1mm}
\end{figure}

To evaluate the impact of the number of blocked flows on the system performance, we plot the number of completed flows and system throughput for the four schemes, which are shown in Figure \ref{fig:QoS_off} and Figure \ref{fig:thro_off}, respectively. In the simulations, the interference threshold $\sigma$ is set to 0.01 and the relay selection parameter $\beta$ is set to 0.53. From the results, we can observe both the number of completed flows and system throughput for all schemes decrease when the number of blocked flow increases. However, compared with MQIS and STDMA, the proposed algorithm always has significantly better performance. This is because when a flow is blocked, we can use a relay with good performance to forward the data,  but MQIS and STDMA don't provide a solution to the blockage problem. Besides, compared with the random relay selection algorithm, our scheme can still maintain higher and more stable performance. This is because when we select the relays, the rate of the relay link and the node contention are considered so that the flows can be completed faster and concurrent transmissions can be fully exploited to improve the transmission efficiency. No matter how many flows are blocked, we can always select proper relays for them. Specifically, when the number of blocked flows equals 10, our scheme improves the number of completed flows by 37.0\% and system throughput by 47.8\% compared with the random relay selection algorithm.
\begin{figure}[tbp]
\vspace*{-2mm}
\begin{minipage}[t]{1\linewidth}
\centering
\includegraphics[width=0.85\columnwidth]{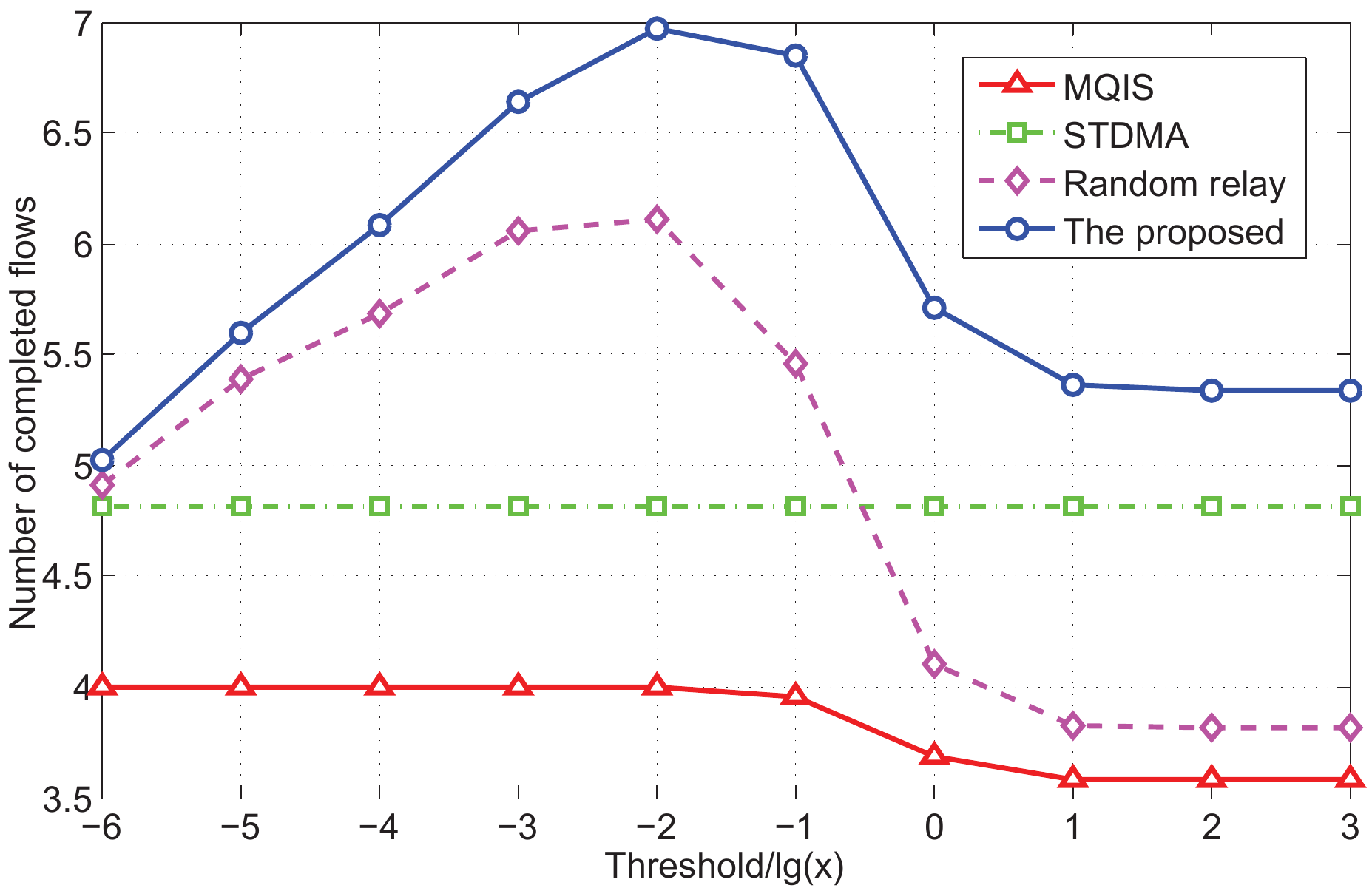}
\end{minipage}%
\vspace*{-3mm}
\caption{Number of completed flows
under different thresholds.}
\label{fig:threshold1}
\vspace*{-0.6mm}
\end{figure}

\begin{figure}[tbp]
\vspace*{-2mm}
\begin{minipage}[t]{1\linewidth}
\centering
\includegraphics[width=0.85\columnwidth]{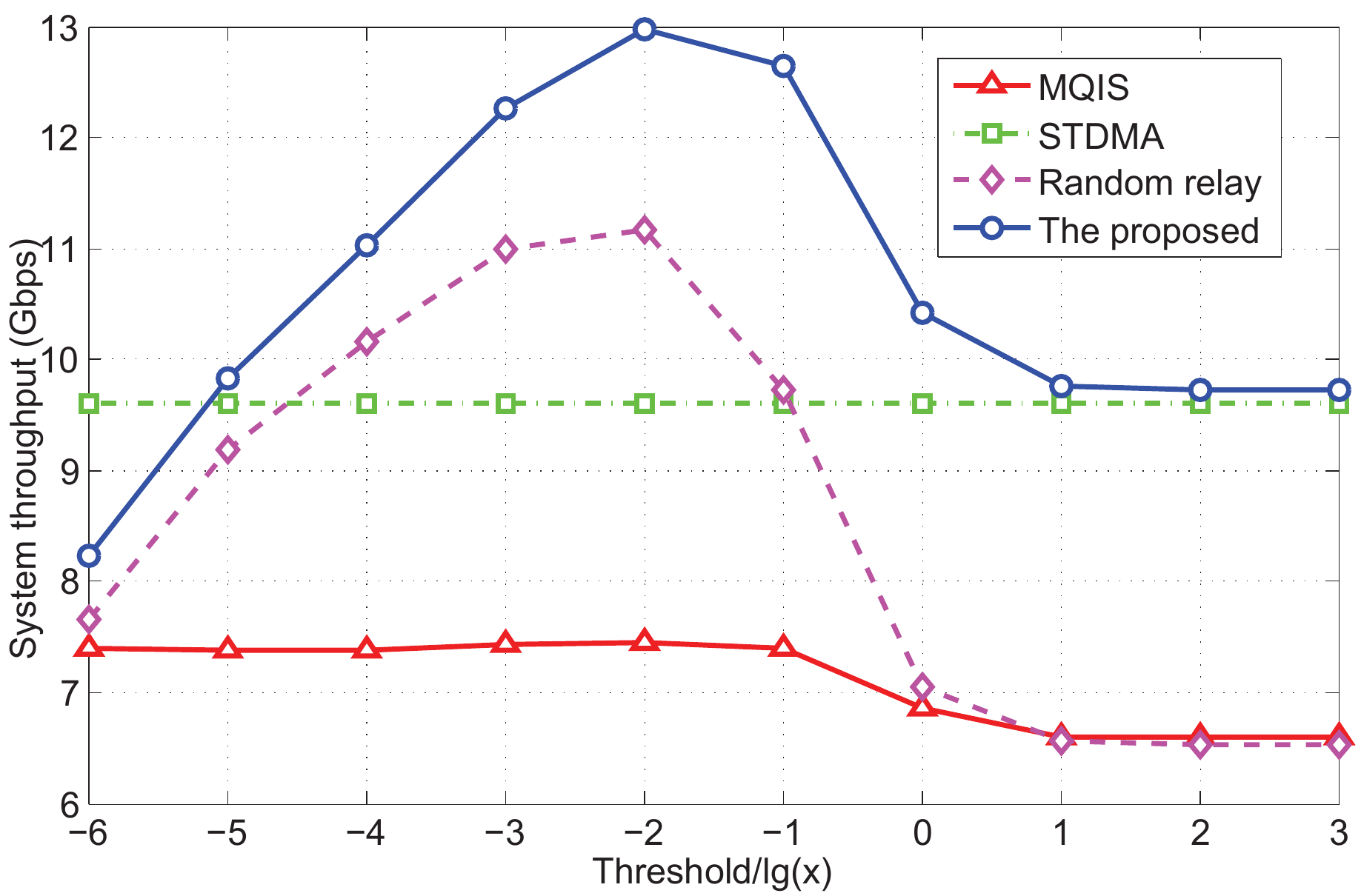}
\end{minipage}%
\vspace*{-3mm}
\caption{System throughput
under different thresholds.}
\label{fig:threshold2}
\vspace*{-1mm}
\end{figure}

In order to investigate the impact of thresholds on the system performance and find the optimal threshold, the two metrics under different interference thresholds are shown in Figure \ref{fig:threshold1} and Figure \ref{fig:threshold2}. Here, the number of blocked flows is set to 5 and $\beta$ is set to 0.53. From the results, we can observe the performance of the proposed algorithm and random algorithm change significantly with the threshold. when the threshold is small, the difference between the two algorithms is negligible. This is because if the threshold is too small, even if the interference between flows is small, they are considered to be in contention and thus concurrent transmissions can't be made full use of. At this time, the threshold is the main limiting factor. However, when the threshold increases, the proposed algorithm could achieve better performance compared with the random relay scheme. This is mainly because we select relays with high link rates, which helps to satisfy the QoS requirements of more flows in the limited time; different blocked flows select different relays, which is beneficial to exploit concurrent transmissions to improve the performance. when $\sigma$ is bigger than $10^{(-2)}$, the performance of these two schemes decreases. This is because that if the threshold is too big, even if the interference between flows is big, they can still be scheduled simultaneously. As a result, the link rates become low and the transmissions become inefficient. When the threshold is bigger than 10, because the interference between flows can't reach to this value, threshold becomes useless and thus the curves of the both algorithms become flat. Therefore, under the simulation conditions in this paper, $\sigma = 10^{(-2)}$ is the optimal threshold. When $\sigma = 10^{(-2)}$, the proposed scheme can improve the number of completed flows by 14.1\% and system throughput by 16.1\% compared with the random relay selection algorithm. Compared with MQIS, which doesn't provide a solution to the blockage problem, the performance of our algorithm always has obvious advantages. As for STDMA, because it doesn't involve the threshold, it doesn't change at all.

\begin{figure}[htbp]
\vspace*{-2mm}
\begin{minipage}[t]{1\linewidth}
\centering
\includegraphics[width=0.85\columnwidth]{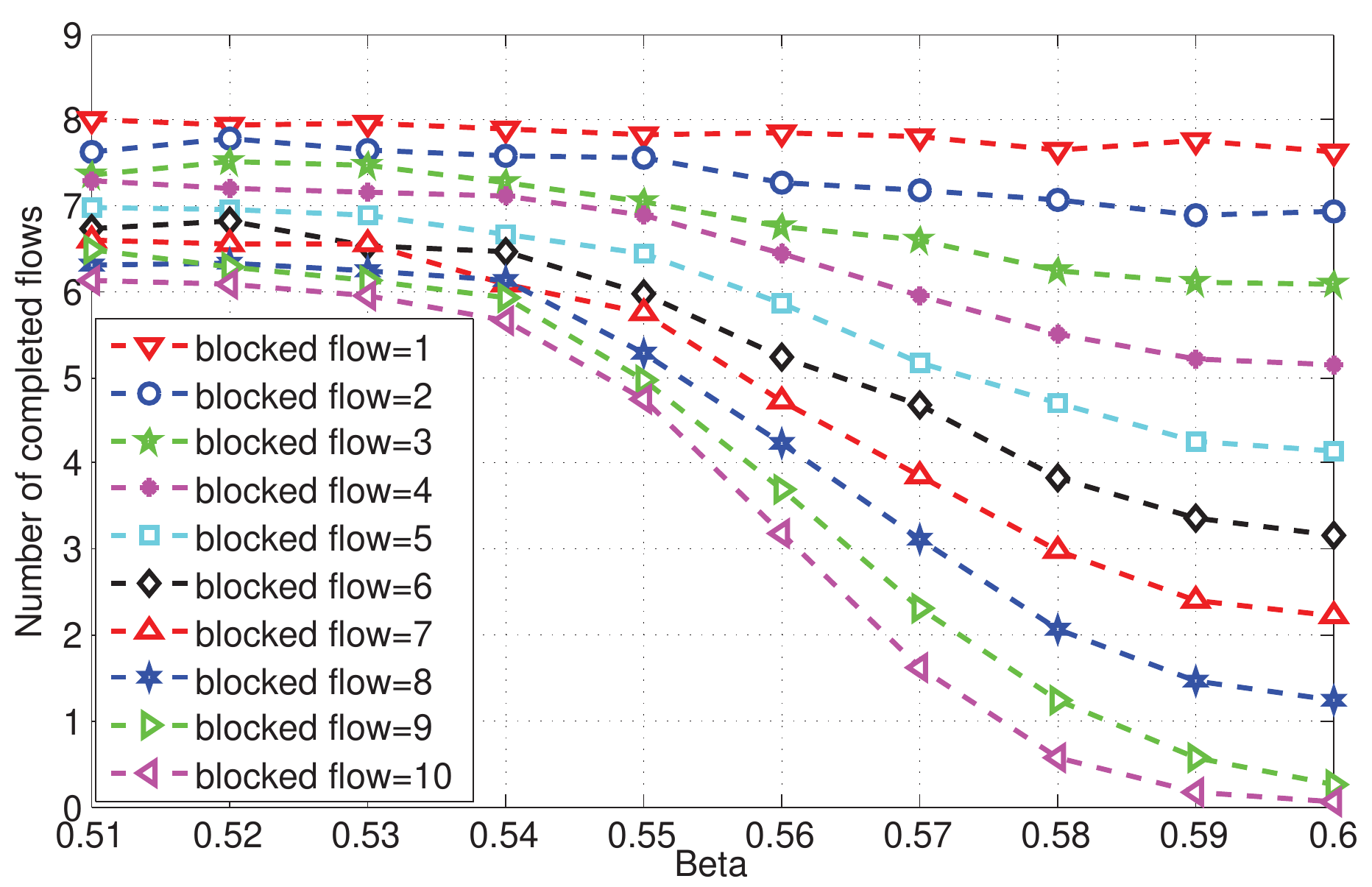}
\end{minipage}%
\vspace*{-3mm}
\caption{Number of completed flows under different relay selection parameters.}
 \label{fig:slv1}
\vspace*{-1mm}
\end{figure}

\begin{figure}[htbp]
\vspace*{-2mm}
\begin{minipage}[t]{1\linewidth}
\centering
\includegraphics[width=0.85\columnwidth]{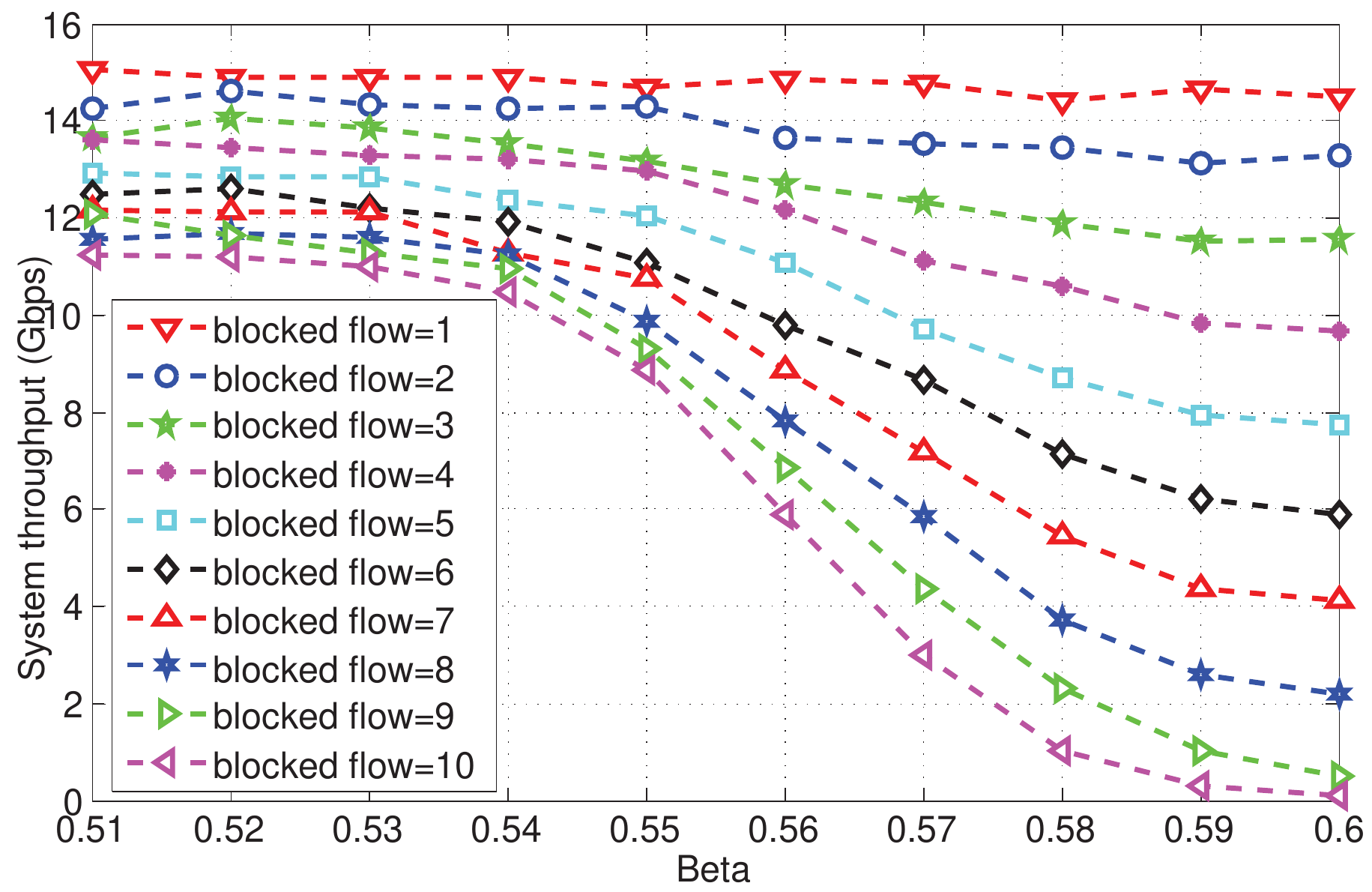}
\end{minipage}%
\vspace*{-3mm}
\caption{System throughput under different relay selection parameters.}
 \label{fig:slv2}
\vspace*{-1mm}
\end{figure}

The impact of relay selection parameter $\beta$ on our protocol is shown in Figure \ref{fig:slv1} and Figure \ref{fig:slv2}. At this time, the threshold is set to 0.01. On one hand, when the number of blocked flows is small, the impact of $\beta$ is not obvious. The greater the number of blocked flows, the greater the effect of $\beta$. On the other hand, the smaller the value of $\beta$, the higher the probability of selecting a relay for the blocked flow(s); so the better the performance. However, when $\beta$ is less than a certain value, the improvement of performance is not obvious. Considering that when $\beta$ is too small, the number of relays in \textbf{Can2} is larger, and choosing the final relay(s) in $\textbf{P}$ is more complex, we should select a proper $\beta$ according to the actual condition. For example, based to our simulation results, when 5 flows are blocked, $\beta = 0.53$ is proper. This is because on one hand, when $\beta = 0.53$, the number of completed flow and the system throughput can maintain a high level, and on the other hand, it's not very complex to find the final relay(s).

\section{Conclusion}\label{S7}
In this paper, we propose a relay-assisted and QoS aware scheduling (RAQS) scheme for the blockage problem in mmWave backhaul networks. First, we propose a relay selection algorithm to forward the data of blocked flow(s), which can select non-repeating relays with high link rates for different blocked flows. Then we propose a heuristic scheduling algorithm to solve the joint scheduling problem of relay paths and backhaul paths, in which both concurrent transmissions and QoS requirements of flows are fully taken into account. The difference between relay path and backhaul path is also considered. Extensive simulations show our scheduling algorithm can effectively overcome the blockage problem, and keep the number of completed flows (i.e., the flows satisfying their QoS requirements in all hops) and system throughput at a high and stable level. In addition, the impact of relay selection parameter is simulated to further guide the relay selection.

In the future work, we will consider other aspects of flows, such as delay, in the problem, and also investigate the delay performance of the
proposed scheme. Besides, we will also investigate the utilization of full duplex technology in mmWave band to improve network performance.

\bibliographystyle{IEEEtran}

\end{document}